\documentstyle[prl,aps]{revtex}

\include{header}

\begin{document}
\draft
\title{Heavy--Fermions in LiV$_2$O$_4$:  Kondo--Compensation vs. Spin--Liquid Behavior?}
\author{H.~Kaps, N.~B\"{u}ttgen, W. Trinkl, A.~Loidl}
\address{Experimentalphysik V, Elektronische Korrelationen und Magnetismus,
Institut f\"{u}r Physik,\\ Universit\"{a}t Augsburg, D--86135
Augsburg, Germany}

\author{M.~Klemm, S.~Horn}
\address{Experimentalphysik II,
Institut f\"{u}r Physik, Universit\"{a}t Augsburg, D--86135
Augsburg, Germany}
\date{\today}
\maketitle

\begin{abstract}
$^{7}$Li~NMR measurements were performed in the metallic spinel
LiV$_2$O$_4$. The temperature dependencies of the line width, the
Knight shift and the spin--lattice relaxation rate were
investigated in the temperature range 30~mK $<$ $T$ $<$ 280~K. For
temperatures $T<1$~K we observe a spin--lattice relaxation rate
which slows down exponentially. The NMR results can be explained
by a spin--liquid behavior and the opening of a spin gap
$\Delta_S$ of the order 0.6~K.

\end{abstract}

Recently, LiV$_2$O$_4$ gained considerable interest after reports
of heavy-fermion formation at low temperatures \cite{Johnston
99,Kondo 97}. Based on heat capacity \cite{Kondo 97},
spin--lattice relaxation \cite{Kondo 97,Mahajan 98}, thermal
expansion \cite{Chmaissem 97} and neutron scattering results
\cite{Krimmel 99} LiV$_2$O$_4$ was treated as a $d$-based
heavy-fermion system \cite{Varma 99}, an interpretation which has
been corroborated by band structure calculations \cite{Anisimov
99}.

LiV$_2$O$_4$ crystallizes in the fcc normal spinel structure and
is characterized by vanadium ions in a $d^1$/$d^2$ mixed valence
state. All band-structure calculations \cite{Anisimov 99,Eyert
99,Matsuno 99,Singh 99} reveal the $t_{2g}$ level close to the
Fermi energy, separated into a lower $A_{1g}$ and a higher $E_g$
orbital. It is emphasized that, in accordance with the
Kondo--lattice interpretation, the $A_{1g}$ electrons ($d^1$,
$S$=1/2) are localized, while the $E_g$ orbitals ($d^{0.5}$)
constitute the band states \cite{Anisimov 99}.

An alternative explanation of the spin--lattice relaxation rates
in LiV$_2$O$_4$ has been provided within the framework of Moriya's
theory \cite{Moriya 79} of ferromagnetic spin fluctuations
\cite{Fujiwara 97} and quasielastic neutron scattering studies
have provided experimental evidence that, in addition to the
unconventional low-temperature properties, LiV$_2$O$_4$ undergoes
a dramatic change of the spin--fluctuation spectrum. Below 40~K
antiferromagnetic fluctuations dominate and the magnetic
relaxation only weakly depends on momentum transfer, while at
higher temperatures ferromagnetic correlations build up and the
relaxation rates linearly increase on momentum transfer as
observed in spin--fluctuation systems.

In this letter we report low-temperature NMR experiments extending
to 30 mK. Temperature dependencies of line width, Knight shift and
spin--lattice relaxation are reported at different measuring
frequencies and external fields from 7.6~MHz/5~kOe  to 137~MHz/
83~kOe, respectively. The temperature dependence of the
spin--lattice relaxation rates convincingly demonstrates that in
LiV$_2$O$_4$ the spin dynamics slows down exponentially but
exhibits no static magnetic order. We conclude that LiV$_2$O$_4$
is similar to other frustrated spin--fluctuation systems like
$\beta$-Mn \cite{Nakamura 97} or Sc doped YMn$_2$ \cite{Ballou
96}, but in addition reveals the opening of a spin gap at low
temperatures. We speculate about an unconventional magnetic ground
state. It is unclear how this behavior can be reconciled with
Kondo--compensation effects.

Polycrystalline samples of  LiV$_2$O$_4$ were prepared by
sintering a mixture of powders of LiVO$_3$ and VO with slight
excess of LiVO$_3$ in order to compensate for Li evaporation.
Platinum crucibles were used for reaction of the powders at 750
${}^\circ$C for 10~days. From EPR and magnetic susceptibility
measurements, we estimated a number of 0.1\% V defects. The NMR
measurements were performed with a phase-coherent pulse
spectrometer and spectra were obtained using field sweeps at
constant frequencies $\omega /2\pi$ = 7.6, 17.3, 72.7 and 137 MHz.
Cryogenic temperatures were provided by a ${}^3$He/${}^4$He
dilution refrigerator with the NMR resonant circuit inside the
mixing chamber. Probing the ${}^7$Li nuclei (spin $I=3/2$,
gyromagnetic ratio $\gamma = 16.546$ MHz/T), we performed
measurements of the line width $\delta$, the Knight shift $K$ and
the spin--lattice relaxation rate $1/T_1$. The spectra were
collected using a conventional $\pi /2 - \tau_D -\pi$ spin--echo
sequence. The line width $\delta$ was deduced from the
field--sweep spectra taking the full width at half maximum. The
spin--lattice relaxation rate $1/T_1$ was determined from the
inversion recovery of the spin--echo intensity. At low
temperatures a stretched exponential relaxation behavior of the
nuclear magnetization was observed as has been reported previously
\cite{Trinkl 99}.

The spin--lattice relaxation rate $1/T_1$ is generally described
via the dynamical susceptibility $Im\chi (q,\omega)$ as
\cite{Narath 68}

\begin{equation}
\label{fluc} {T_1}^{-1}=\frac{\gamma_n^2k_BT}{2\mu_B^2}
{A_{\text{hf}}}^2 \sum_q \frac{Im\chi(q,\omega,T)}{\omega}.
\end{equation}

Here $A_{\text{hf}}$ is the hyperfine coupling which is assumed to
be isotropic and temperature independent. A Lorentzian-type
function is used for the frequency dependence of the dynamical
susceptibility

 \begin{equation}
\label{ansatz} Im\chi (q,\omega) = \chi (q)\cdot
\frac{\omega\Gamma}{\omega^2+\Gamma^2} \;\: .
\end{equation}

where $\Gamma$ is the magnetic relaxation rate which measures the
characteristic energy of the spin--fluctuation spectrum. The
prefactor $\chi(q)$ denotes the wave-vector dependent static
susceptibility. In a first approach we neglect any $q$ dependence
of the static susceptibility $\chi_0$.

Under the assumptions outlined above and including a 'metallic'
Korringa term, the spin--lattice relaxation rate is given by
\begin{equation}
{T_1}^{-1}= a\cdot T + b\;T\;\chi_0\cdot\frac{\Gamma
(T)}{\omega_0^2+\Gamma^2(T)} \label{eq3}
\end{equation}

where $b=\gamma_n^2k_BT{A_{\text{hf}}}^2/2\mu_B^2$ is a constant
and $\omega_0$ gives the Larmor frequency of the NMR experiment.
The first term takes Korringa-type relaxations into account which
stem from contributions of the band states. Using Eq.(\ref{eq3})
the characteristic temperature dependence of the spin--lattice
relaxation rate of heavy-fermion systems can be recovered assuming
a Curie-Weiss like susceptibility, $\chi_0=C/(T+\alpha T^*)$ with
$\alpha= \sqrt{2}$ and the characteristic Kondo temperature $T$*
\cite{Bernal 95} and a magnetic relaxation rate that reveals a
temperature dependence as $\Gamma (T)=\Gamma_0 + \beta \sqrt{T}$
\cite{Cox 85}. Under these simple assumptions the spin--lattice
relaxation rate $1/T_1$ reveals a cusp close to the characteristic
temperature $T$* and a Korringa behavior with a highly enhanced
slope for $T \ll T^*$. These are characteristic features which are
observed experimentally in heavy-fermion compounds \cite{Asayama
88}. In LiV$_2$O$_4$ the temperature dependence of $1/T_1(T)$
nicely resembles this behavior \cite{Kondo 97,Mahajan 98,Fujiwara
97,Trinkl 99} and the square root dependence of the magnetic
relaxation rate has been proven by neutron scattering experiments
by Krimmel {\em et al.}\cite{Krimmel 99}.

Figure 1a shows the temperature dependence of the Knight shift at
two measuring frequencies. The Knight shift provides a direct
measure of the local static susceptibility. The cusp-like shape at
approximately 30~K indicates the characteristic temperature $T^*$.
The Knight shift is frequency/magnetic-field independent above
$T^*$, but a significant dependence evolves below the temperature
of the cusp maximum. For $T<T^*$ the Knight shift decreases
approximately logarithmically and levels off at a constant value
below 0.3~K. Figure 1b shows the temperature dependence of the
line width $\delta$, again at two measuring frequencies. For both
frequencies the line width continuously increases on decreasing
temperatures and saturates below 1~K. For the higher frequency and
field the increase significantly is enhanced. The line width is
dominated by an inhomogenous broadening due to local magnetic
fields \cite{Mahajan 98,Trinkl 99}. The constant value at low
temperatures signals that this internal fields become frozen and
remain constant on the time scale of the experiment. However,
below 0.1~K only a small fraction of nuclear spins contribute to
the signal as it is clearly shown by the drastic decrease of the
NMR intensity towards the lowest temperatures (see the inset in
Fig.1).

The temperature dependence of the spin--lattice relaxation rate is
shown in Fig. 2. Again, the maximum close to 30 K indicates the
characteristic temperature $T^*$. This behavior nicely resembles
the results found by Kondo {\em et al.}\cite{Kondo 97}. A
Korringa-like behavior which has been determined by these authors
in a temperature range from 1.5 K $<T<$ 6 K is indicated as solid
line in Fig. 2. Our low temperature values of the spin--lattice
relaxation rate at 72~MHz remain slightly enhanced. This may be
due to the fact that we analyzed the data assuming a stretched
exponential recovery of the relaxation, as a pure exponential fit
did not work especially at temperatures below 1~K \cite{Trinkl
99}. Astonishingly, at low measuring frequencies a clear
cusp-shaped maximum appears at approximately 0.6~K which becomes
almost suppressed at higher frequencies.

To study the anomalous low--temperature relaxation in more detail,
we performed a series of experiments at different measuring
frequencies and associated magnetic fields (Fig. 3a). Below 2~K
the temperature dependence of $1/T_1$ reveals a significant
frequency dependence and the nuclear relaxation is strongly
enhanced at low frequencies. This behavior clearly reveals a
similar characteristic as the Li nuclear relaxation observed in Li
doped CuO and NiO by Rigamonti and coworkers \cite{Rigamonti 97}
which has been compared to the spin dynamics in cuprate
superconductors. On the basis of this interpretation the cusp-like
anomalies in Fig. 3a below 1~K signal the slowing down of spin
fluctuations on a time scale given by the NMR experiments. We
easily can incorporate this phenomenon in our model for correlated
materials. Under the assumption of an exponentially increasing
magnetic relaxation $\tilde{\Gamma} = \Gamma(T) \cdot
exp(-\Delta/k_BT)$ the spin--lattice relaxation rate $1/T_1$ can
be calculated using Eq.(\ref{eq3}). Below 1~K the magnetic
relaxation rate $\Gamma(T)$ is dominated by the exponential
decrease, but it recovers the square root dependence at elevated
temperatures. From the results of these calculations we obtain a
rough estimate of the spin gap $\Delta$ to be of the order 1~K
(Fig. 3b). The qualitative agreement is surprising having the very
restrictive model assumptions in mind and especially neglecting
the influence of the external magnetic field. On the basis of this
results we conclude that, on decreasing temperatures and well
below 1~K the spin fluctuations slow down exponentially and at the
cusp maximum the measuring frequency directly corresponds to the
relaxation rate. At temperatures $T <$ 0.1~K the slow relaxation
regime ($\Gamma \ll \omega$) is reached. This observation of a
slow spin dynamics down to the lowest temperature is in agreement
with the $\mu$-SR results where the sample with the lowest
impurity concentration revealed a slowing down of spin
fluctuations with no signature of static freezing \cite{Kondo
97,Merrin 98}.\\

In Fig. 3c we show the low temperature spin--lattice relaxation
rate in an Arrhenius representation to indicate the exponential
increase of $1/T_1(T)$ at low temperatures. We have subtracted the
limiting low--temperature spin--lattice relaxation which is
dominated by defect spins (see Inset in Fig. 1). For all
frequencies we find an activated (solid lines) behavior
corresponding to an effective spin gap of $\Delta_S$=0.6~K. The
frequency dependence of $1/T_1$ is weaker than $1/\omega^{2}_0$
which is expected for the slow relaxation regime (see Eq.(3)).
This fact points towards a broad distribution of relaxation rates
\cite{Rigamonti 97}.\\

How can these results be reconciled with a heavy-fermion picture
and what is the intrinsic ground state of LiV$_2$O$_4$. Of course
LiV$_2$O$_4$ is close to magnetic order. For example 5\% Zn doping
induces spin--glass freezing at approximately 2.5~K \cite{Trinkl
99}. Certainly the same is true for prototypical heavy-fermion
systems and we would like to recall that, depending on the exact
Ce stoichiometry, CeCu$_2$Si$_2$ reveals superconductivity
(S-phase) or magnetic order (A-phase). The regime of long--range
magnetic order and $d$--wave superconductivity is separated by a
phase which is dominated by slow magnetic fluctuations
\cite{Ishida 99}. Based on a variety of different experiments,
also the small--moment magnetism in UPt$_3$ most probably is
dynamic in origin. An oscillating spin--density wave, with a
characteristic fluctuation rate of some GHz, has been proposed to
explain the results in this system \cite{Okuno 98}.\\

All these facts reveal striking similarities with the experimental
observations in LiV$_2$O$_4$. However, LiV$_2$O$_4$ also should be
compared to canonical spin liquids. It is a frustrated magnet with
the V ions forming  a lattice of corner sharing tetrahedra. The
Sommerfeld coefficient reaches 420~mJ/(mol K$^2$) \cite{Kondo 97}
and the low-temperature magnetic relaxation rate is 0.5~meV
\cite{Krimmel 99}. The spin liquid Sc$_{0.03}$Y$_{0.97}$Mn$_2$
reveals the same structural frustration, has a specific heat
coefficient of 160 mJ/(mol K$^2$) and a magnetic relaxation rate
of 8~meV. It's striking heavy-fermion like behavior has clearly
been addressed \cite{Ballou 96}. $\beta$-Mn is another
geometrically frustrated magnet. Also this system reveals no
static magnetic order and is characterized by a Sommerfeld
coefficient $\gamma$ = 70 mJ/(mol K$^2$) and a weakly temperature
dependent magnetic relaxation rate $\Gamma \approx$ 20~meV
\cite{Nakamura 97}.\\

Astonishingly, spin liquids and heavy-fermions reveal very similar
dynamical susceptibilities. In both cases low-lying and usually
gapless magnetic excitations govern the dynamic susceptibility.
However, the underlying physics seems to be rather different: In
spin liquids long-range magnetic order is suppressed by
topological magnetic frustration, soft magnetic excitations are
enhanced favoring the formation of local singlets and the
existence of an unconventional non-N\'{e}el magnetic ground state
\cite{Chandra 91}. In heavy--fermion systems Kondo compensation
yields an enhanced density of states at the Fermi energy driving
the formation of heavy quasiparticles. Antiferromagnetic spin
fluctuations screen the local moments.\\

What type of exotic ground state is established in LiV$_2$O$_4$?
At temperatures $T>$1~K the dynamical susceptibility is
characteristic of a strongly correlated electron system. It is in
this regime where spin--liquids and heavy--fermion systems behave
similar. However, our results show that the magnetic relaxation
$\Gamma (T)$ and consequently the spin--lattice relaxation rate
$1/T_1$ slow down exponentially indicating the opening of a spin
gap of the order 0.6~K. The gap could be due to dynamic singlet
pairing. This interpretation is in accord with the 'cooperative
paramagnet' which has been proposed by Villain \cite{Villain 79}
as a possible ground state of cubic spinels: The spins of each
tetrahedron form antiparallel pairs, at least on a time scale
large compared to the inverse NMR frequencies.\\

We believe that LiV$_2$O$_4$ is dominated by frustration effects
rather than by moment compensation. Can we exclude a spin--glass
transition? Exotic spin--glass behavior in systems without
disorder has been proposed by Villain \cite{Villain 79} who
pointed out that canonical spin--glass behavior is unlikely to
occure for spinels like LiV$_2$O$_4$. In addition, taking a
Kondo-lattice temperature of 30~K, it is hard to understand how
random freezing of moments can appear well below 1~K with fully
compensated moments. Also from a purely experimental point of
view, the (dynamic) transition below 1~K behaves significantly
different as compared to the spin--glass transitions observed in
disordered Li$_{1-x}$Zn$_x$V$_2$O$_4$ with Zn concentrations $x>
0.05$ \cite{Trinkl 99}. Instead we believe to have observed
singlet formation (or another complex non-N\'{e}el state) in
topologically frustrated LiV$_2$O$_4$ and further experiments to
elucidate the ground state properties are highly needed.\\

This work has partly been supported by the Sonderforschungsbereich
484 of the Deutsche Forschungsgemeinschaft and the BMBF under the
contract number EKM 13N6917/0.

\begin{figure}
\caption{a) ${}^7$Li Knight shift $K$ vs. temperature $T$ in
 LiV$_2$O$_4$ at two measuring frequencies/external fields, respectively:
 ($\circ$) 17.3~MHz/10~kOe and ($\Box$) 72.7~MHz/44~kOe.
 b) ${}^7$Li line width $\delta$ vs. temperature $T$.  Inset:
 ${}^7$Li-NMR intesity multiplied by temperature and normalized $I\times T$ at $T$=1~K.}
\label{f1}
\end{figure}

\begin{figure}
\caption{Semi-logarithmic representation of the
 ${}^7$Li spin--lattice relaxation rate $1/T_1$ vs. temperature $T$ in
 LiV$_2$O$_4$ at two measuring frequencies/external fields, respectively:
 ($\circ$) 17.3~MHz/10~kOe and ($\Box$) 72.7~MHz/44~kOe. The solid line
 indicates a Korringa relation which was found by
  Kondo {\em et al.}\protect\cite{Kondo 97}}
\label{f2}
\end{figure}

\begin{figure}
\caption{a) ${}^7$Li spin--lattice relaxation rate $1/T_1$ vs.
temperature $T$ in LiV$_2$O$_4$. b) Model calculations with one
unique set of parameters as described in the text
 ($a$ = 2.54 s$^{-1}$K$^{-1}$, $b$ = 5.34 $\cdot 10^{11}$s$^{-2}$, $\Gamma_0$ = 1.68 GHz,
 and $\Delta$ = 1.47 K). c) Arrhenius plot of the spin--lattice relaxation rate $1/T_1$. The
 solid lines indicate an activated behavior using a spin gap $\Delta_S$ = 0.6~K.}
\label{f3}
\end{figure}

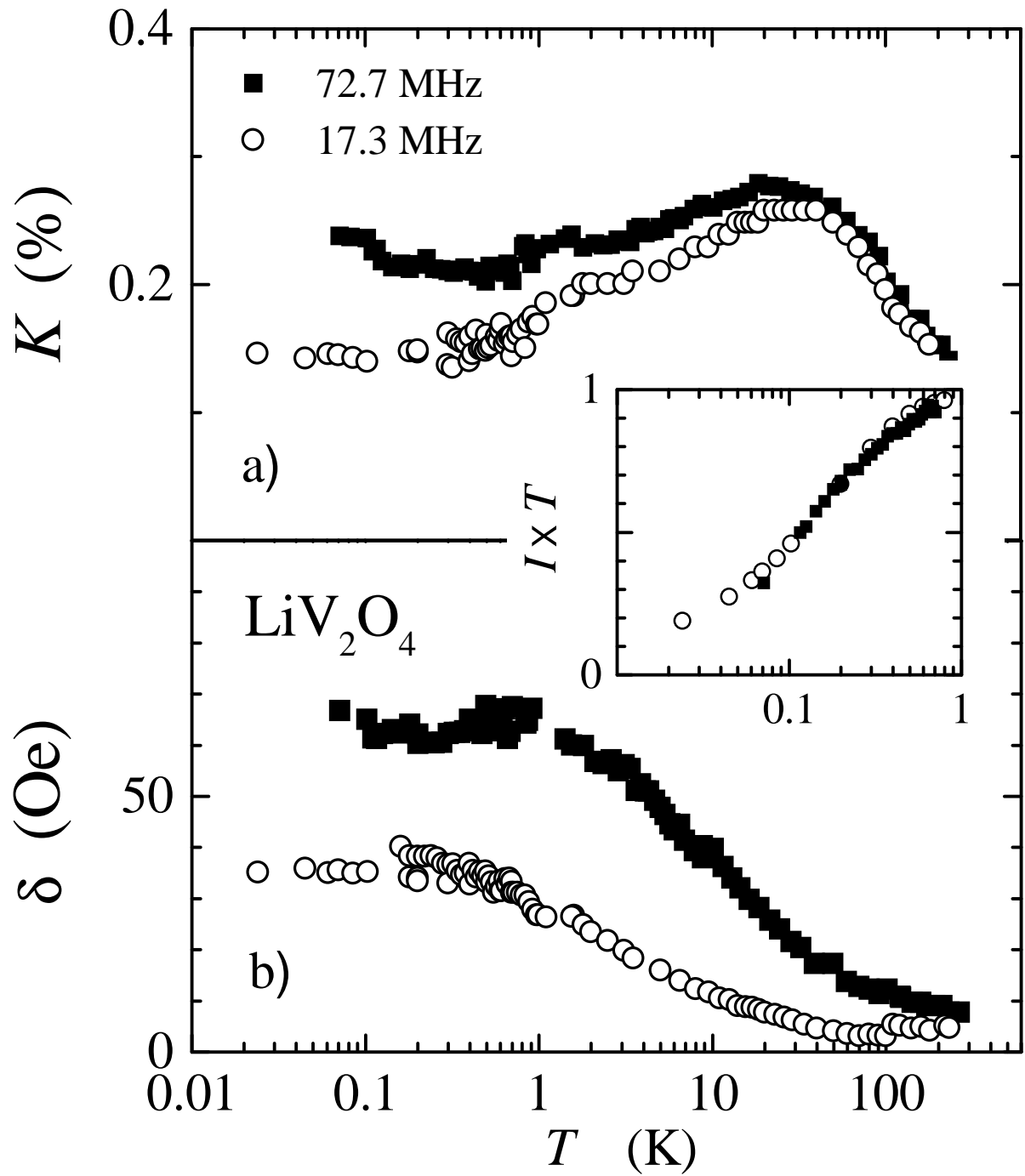
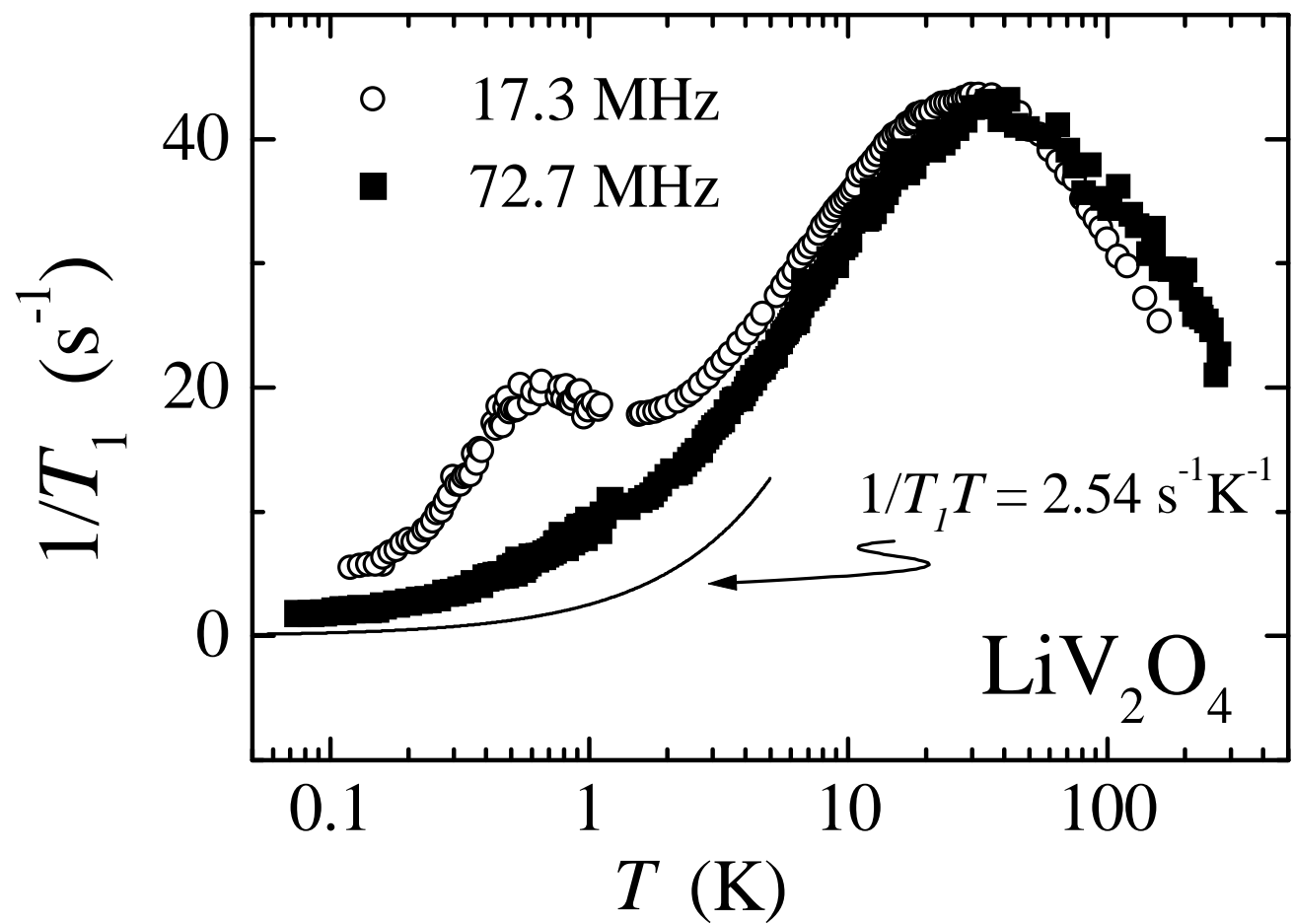
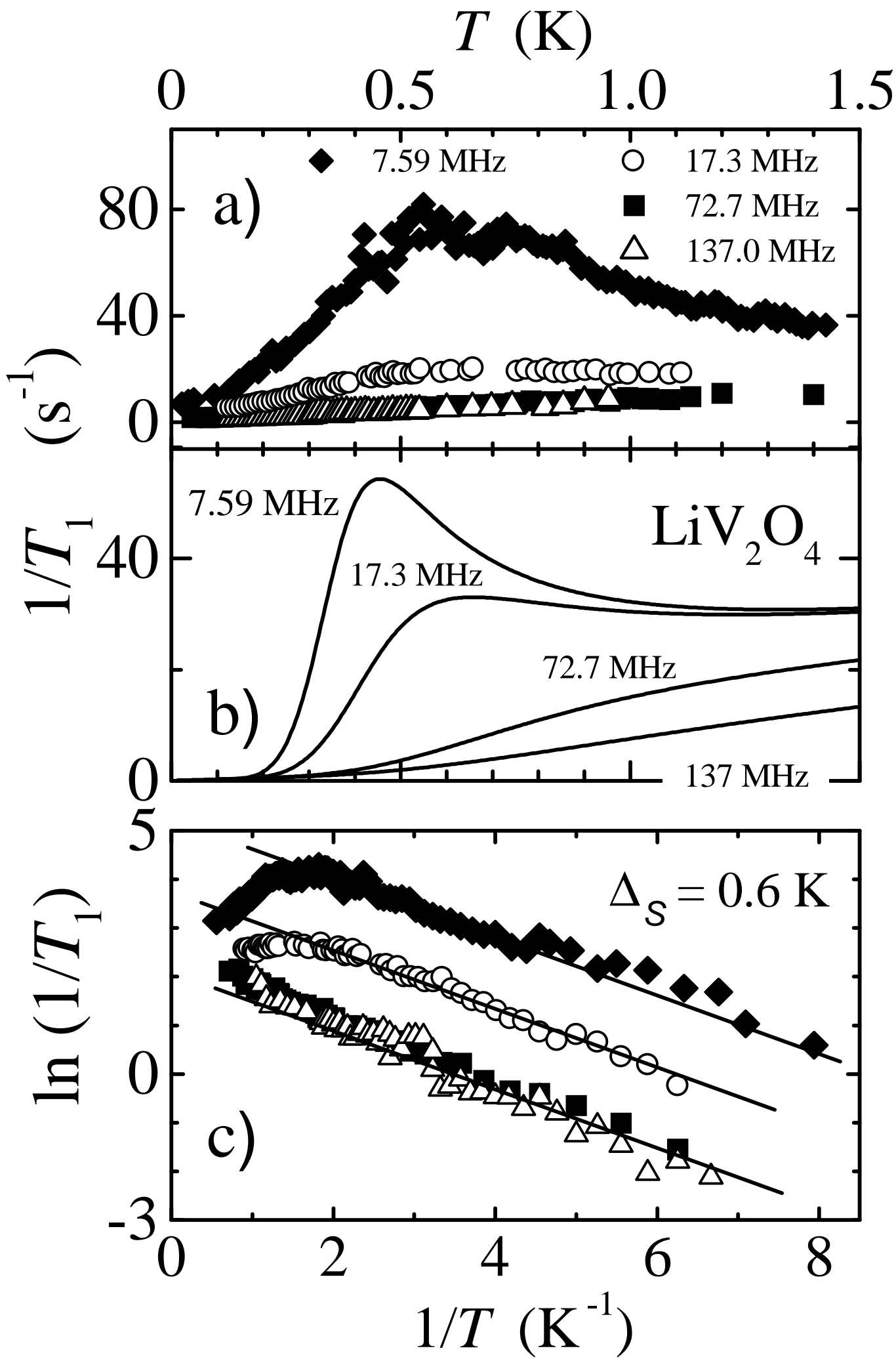
\end{document}